\documentclass{article}
\usepackage{amsmath}
  \usepackage{graphics} 
  \usepackage{epsfig} 
 \usepackage[colorlinks=true]{hyperref}
 \hypersetup{urlcolor=blue, citecolor=red}

\newtheorem{theorem}{Theorem}[section]
\newtheorem{corollary}{Corollary}

\newtheorem{proposition}{Proposition}

\newtheorem{definition}[theorem]{Definition}
\newtheorem{remark}{Remark}

\usepackage{psfrag}

\def\dem {\noindent {\bf Proof: }}

\def\sqw{\hbox{\rlap{\leavevmode\raise.3ex\hbox{$\sqcap$}}$%
\sqcup$}}
\def\findem{\ifmmode\sqw\else{\ifhmode\unskip\fi\nobreak\hfil
\penalty50\hskip1em\null\nobreak\hfil\sqw
\parfillskip=0pt\finalhyphendemerits=0\endgraf}\fi}

\newcommand{\Z}{\mathbb Z}
\newcommand{\F}{\mathcal F}
\usepackage{amsmath}
\usepackage[psamsfonts]{amssymb}
\usepackage[psamsfonts]{eucal}

\def \setR {\mathbb{R}}
\def \setZ {\mathbb{Z}}
\def \Z {\mathbb{Z}}
\def \setQ {\mathbb{Q}}
\def \setN {\mathbb{N}}
 \def \<#1,#2> {{\left\langle #1\mathbin,#2\right\rangle}}
 \def \S {\mathcal{S}}
\def \X(#1,#2)   { X_{#1 #2} }
\def \(#1,#2) {{\left(#1\mathbin,#2\right)}}
\def \norme #1 {\left \| #1 \right \|}
\def \one  {\hbox{\rm I}\hskip -3pt 1}
\def \virg {\, , \;}
\def \abs #1  {{\left |#1\right|}}
\def \ZZ  {\setZ_2}
\def \QQ  {{\setQ_2}}

\begin{document}

\title{A 2-adic approach of the human respiratory tree}
\date{April 30, 2010}

\maketitle

\centerline{\scshape Fr\'ed\'eric Bernicot}
\medskip
{\footnotesize
 \centerline{Laboratoire Paul Painlev\'e - CNRS, Universit\'e Lille 1,}
 \centerline{ 59655 Villeneuve d'Ascq Cedex, France}
}  %

\medskip

\centerline{\scshape Bertrand Maury }
\medskip
{\footnotesize
 \centerline{Laboratoire de Math\'ematiques, Universit\'e Paris-Sud 11,}
   \centerline{91405 Orsay Cedex, France}
} 
\medskip

\centerline{\scshape Delphine Salort }
\medskip
{\footnotesize
 \centerline{ Institut Jacques Monod, Universit\'e Paris Diderot}
   \centerline{B{\^a}t. Buffon, 15 rue H{\'e}l{\`e}ne Brion, 75013 Paris, France
}
} %

\bigskip

\begin{abstract}
We propose here a general framework to address the question of trace operators on a dyadic tree. This work is motivated by the modeling of the human bronchial tree which, thanks to its regularity, can be extrapolated in a natural way to an infinite  resistive tree. The space of  pressure fields at bifurcation nodes of this infinite  tree can be endowed with a Sobolev space structure, with a semi-norm which measures the instantaneous rate  of dissipated energy. We aim at describing the behaviour of finite energy pressure fields  near the end.
The core of the present approach is an identification of 
the set of ends with the ring $\ZZ$ of 2-adic integers. Sobolev spaces over $\ZZ$ can be defined in a very natural way by means of Fourier transform, which allows us to establish precised trace theorems which are formally quite similar to those in standard Sobolev spaces, with a Sobolev regularity which depends on the growth rate of resistances, i.e. on geometrical properties of the tree. Furthermore, we exhibit an explicit expression of the ``ventilation operator'', which maps pressure fields at the end of the tree onto fluxes, in the form of a convolution by a Riesz kernel based on the 2-adic distance.

\end{abstract}

\maketitle

\section{Introduction, modelling aspects}

The  human bronchial tree  can be seen as a set of dyadically connected pipes, which sums up to 23 bifurcation levels from the trachea to terminal branches, on which gas exchanges occur. 
Thanks to Poiseuille's law for a pipe, which states a proportionalty  relation between  air flow rate and  pressure jump, it behaves as a fluid conductor like a resistive network, where pressure at bifurcating nodes plays the role of the electric potential, and air flow the role of electric intensity. 
Schematically\footnotemark, air is driven to the alveoli (zone of gas exchange with the blood) by a negative pressure maintained on the outlets during inspiration. 
In the situation where pressure is a constant $P$  and the tree is regular (i.e. resistances are the same for all pipes of a given generation), the overall process follows a generalized Poiseuille law (fluid counterpart of  Ohm's law)
$$
\hbox{(atmospheric pressure)} - P  = R \times  \hbox{(flux)},
$$
where $R$ is  the global resistance.
Yet, as soon as pressures are not uniform, or if the tree is no longer regular (which can happen for example during an asthma crisis), this scalar  ohmic law has be to extended. For the  idealized tree with $2^{23}$ outlets, this law takes the form of a linear (if atmospheric pressure is set at $0$) relation between a collection of $2^{23}$ values for the pressure, and a vector of fluxes.  The corresponding matrix $R$ can be written explicitly as a function of the resistances (see~\cite{GMM}).

\footnotetext{The reality is a bit more complex, as gas exchanges take place earlier in the tree, from the $16^{th}$ generation to the last one.}

In a recent paper~\cite{msv}, a model of the respiratory system as an infinite tree was proposed. 
This extrapolation of the finite resistive tree to an infinite one is 
natural  because the actual bronchial tree exhibits some geometric regularity. Indeed, it is shown in~\cite{weibelgeom} that dimensions of the pipes (at least in the conducting, central part of the tree, i.e. between generations 6 and 17) progress in a geometric way, with a reduction factor $\lambda$ close to $0.85$. As the Poiseuille resistance of a pipe scales like $\ell/r^4$ (where $\ell$ is the length and $r$ the radius), resistance at  generation  $n$ scales like $\alpha^n$, with $\alpha = 1/\lambda ^3 \approx 1.63$. For the extrapolated infinite tree, as  generation $n$  contains  $2^n$ resistances in parallel, one obtains a finite global resistance
$$
R = \sum \left ( \frac {\alpha} {2 }\right ) ^n < +\infty, 
$$
and this finiteness is stable under small  perturbations of the geometry.
Next step  consists in replacing  the collection of discrete pressures at the end of the actual tree by a ``continuous'' field (a function defined over the uncountable set of ends).
A first  way to identify the corresponding  trace space is immediate. Denoting by $V$ the set of vertices of  tree $T$, $p(x)$ the pressure at vertex $x$,  writing $x\sim y$ for connected vertices and $r(x,y)$ the resistance of the corresponding edge, Sobolev space $H^1(T)$ over $T$ is defined as all those pressure fields such that 
$$
\sum _{x\sim y } \frac 1 {r(x,y)} \abs { p(y) - p(x)} ^2 < \infty.
$$
Note that  the previous definition makes sense for any kind of network with bounded connectivity~: the quantity above simply represents the instantaneous rate of dissipated energy by viscous effects. 
Denoting  by $H^1_0$ the closure of finitely supported fields, a first expression of the trace space is the quotient  $H^1 / H^1_0$. As detailed in~\cite{soardi} (in a probabilistic framework), non triviality of this space, which is equivalent to $R < +\infty$, indicates that ``something happens''  at infinity, i.e. it does make sense to prescribe certain pressure fields at infinity to drive some fluid through the tree.

In order to obtain a more explicit description of the space of trace functions (pressure fields on the set of ends), an explicit construction of a Hilbert basis of  harmonic, finite energy functions is proposed in~\cite{msv}. Such functions are in some sense piecewise constant at infinity, so that their trace can be defined canonically. General trace theorems then follow in a standard way  by density.

Following  a suggestion by P. Colmez, 
we propose here an alternative approach, based on the identification of the boundary of the tree (set of ends) to the ring of 2-adic integers $\setZ_2$, and a different strategy to define traces.  To any pressure  field $p$ defined over the tree, we shall consider its restriction $p_n$  to the $n$-th generation, consider the corresponding function $\tilde p_n$ in the Schwartz space of $\ZZ$  (constant over each bunch of leafs stemming from any of the $n$-th generation vertices), and show that this sequence converges in $L^2(\ZZ)$, and possibly in a stronger way  under some condition on the resistance growth. Regularity of the corresponding trace function will be expressed in terms of behaviour near infinity of the Fourier transform (Sobolev-like regularity).

The approach presented here may seem excessively and artificially abstract considering the actual  object it aims at reproducing in some way. Indeed, we are aware that some  parts  of what we present can be (and actually have been in some cases) presented in a different  and less abstract setting, in particular 
\begin{enumerate}
\item[$(i)$] Regularity  of the trace functions; 

\item[$(ii)$] Expression of the Neuman-Dirichlet  operator, which maps flow fields onto pressure fields.

\end{enumerate}

Concerning  $(i)$,   functions over the set of ends of the tree can be described  in terms of  $A^s$ regularity, as presented in~\cite{msv}, without any reference to 2-adic integers. 
Yet, as we will see, the $\setZ_2$ approach allows a very natural and concise way to define  Sobolev spaces in the Fourier setting, which enlights the deep similarities with trace operators 
in the Partial Differential Equations context.
As for the Neuman Dirichlet  operator, it  can be represented by matrices for finite tree (see~\cite{MMSV}), and by kernel
 operators for infinite tree (see~\cite{GMM}). We will present here how the $\setZ_2$ approach  makes it possible to express them as a convolution by a Riesz kernel with
 a exponent directly related to the geometric growth of  resistances.

This paper is structured as follows :  in Section~\ref{sec:ident}, we present  the identification between the set of ends of a dyadic tree and $\ZZ$, 
 and  we give a first trace theorem based on this identification. In Section~\ref{sec:sob} we give some regularity properties of trace functions (Sobolev regularity). 
Section~\ref{sec:DN} is dedicated to Dirichlet- Neuman and Neuman-Dirichlet operators, and in Section~\ref{sec:embed} we investigate the possibility to imbed the end of the 
infinite tree onto a domain of $\setR^d$ (actual domain occupied by a real lung). Finaly we gather in the appendices some  facts on Fourier Analysis  and Sobolev spaces  on $\setZ_2$.

In what follows the ideal dyadic infinite  geometric tree, with resistances following a geometric growth in $\alpha^n$ (with $\alpha$ close to $1.63$ for a healthy lung, as stated before) will play a central role. 
Yet, we shall present results with  maximal generality, allowing when it is  possible non-regular trees (non-uniform resistances   within a generation). 

\section{Identification with $\setZ_2$}
\label{sec:ident}

We first gather some definitions and standard properties
of 2-adic numbers (see e.g.~\cite{colmez}).
 
 For any $z \in  \setZ$, $z = z' 2^\alpha$, with $z'$ odd, one defines valuation $v_2(z)$ as $ \alpha$.
 One extends this definition to rational numbers by setting $v_2(q) = v_2(a) - v_2(b)$ for any 
  $q = a/b\in  \setQ$, $q \neq 0$, and $v_2(0) = +\infty$. Now setting $\abs {q } _2 = 2^{-v_2(q)}$, the 2-adic distance over $\setQ$ is defined as
  $$
  (q,q') \in \setQ \times \setQ \longmapsto  \abs { q'-q} _2.
  $$
  This distance is ultrametric~: it verifies a strong triangle inequality
  $$
  \abs { q'' - q } _2 \leq \max\left (  \abs { q'' - q' } _2,   \abs { q' - q } _2\right ) . 
  $$
  As a consequence, a ball is centered at any of its elements.
The complete closure of $\setQ$ for this distance  is  called $\setQ_2$.
 Any element of $\setQ_2$ can be identified to a  series
 \begin{equation}
 \label{eq;rep}
q =  \sum_{n = k}^{+\infty}  a_n 2^{-n}, 
\end{equation}
where $k\in \setZ$, $a_n \in \{0,1\}$, and therefore  written (here in the case $k < 0$) as 
$$
q = \dots a_n \dots a_1 a_0, a_{-1} a_{-2}\dots a_{k}.
$$
The ring $\setZ_2$ of 2-adic integers is defined as the closed unit ball of $\setQ_2$. In the above representation, it corresponds to the case where $a_n=0$ for all $n<0$.
 
\begin{figure}
\psfragscanon
\psfrag{0}{$0 $}
\psfrag{1}{$1 $}
\psfrag{2}{$2 $}
\psfrag{3}{$3 $}
\psfrag{4}{$4 $}
\psfrag{5}{$5 $}
\psfrag{6}{$6 $}
\psfrag{7}{$7 $}
\psfrag{8}{$8 $}
\psfrag{9}{$9 $}
\psfrag{10}{$10 $}
\psfrag{11}{$11 $}
\psfrag{12}{$12 $}
\psfrag{13}{$13 $}
\psfrag{14}{$14 $}
\psfrag{15}{$15 $}
\psfrag{Z2Z}{$\setZ/2\setZ$}
\psfrag{Z4Z}{$\setZ/4\setZ$}
\psfrag{Z8Z}{$\setZ/8\setZ$}
\psfrag{Z16Z}{$\setZ/16\setZ$}
\centerline{
\includegraphics[width=1\textwidth]{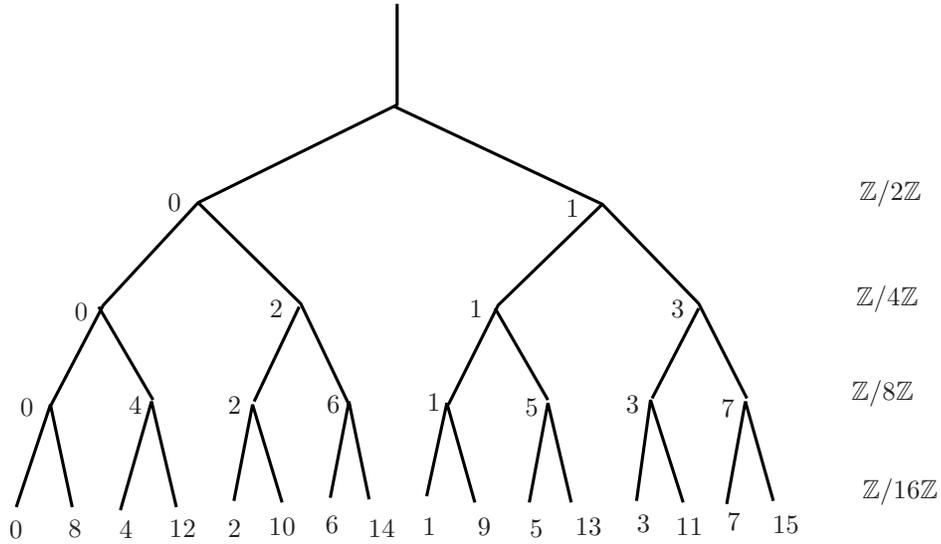}
}
\label{fig:tree}
\caption{Dyadic tree}
\end{figure}

Let now  explain how the set of ends $\partial T $ of 
 a  dyadic tree $T$  can be identified with $\setZ_2$. 
The set of vertices can be seen as the 
disjoint union of the $\setZ/ 2^n\setZ$'s, for $n=0$, $1$, $\dots$,
as illustrated by Fig.~\ref{fig:tree}.
We shall denote by $x_n^k$ the vertex $k$ at generation $n$ 
(i.e.  $x_n^k$ is $k$ considered  as an element of $\setZ/ 2^n\setZ$).
Now denoting by $\varphi _n^m$, with $n< m$,  
the  canonical surjection from $\setZ/ 2^m\setZ$ onto  $\setZ/ 2^n\setZ$, 
the set of edges of $T$, namely $E$,  consists of all those couples 
$$
( x_n^k,x_{n+1}^\ell ) \in  \setZ/ 2^n\setZ \times \setZ/ 2^{n+1}\setZ
$$
such that $x_n^k = \varphi _n^{n+1} ( x_{n+1}^\ell)$. The corresponding edge is denoted by $e_{n+1}^\ell$.

The set $\partial T$ of ends of $T$  (infinite paths toward infinity)
can be represented by the projective limit of the system 
$\left (\setZ/2^n\setZ,\varphi _n^{m}\right )$~:
$$
\partial T:=\varprojlim \left (\setZ/2^n\setZ,\varphi _n^{m}\right )  = 
\left \{ \left (z_n\right )_{n\in\setN} \in \Pi \left (\setZ/2^n\setZ
 \right ) \virg \varphi_n^{n+1} (z_{n+1}) = z_n\quad \forall n\geq 0
\right \}.
$$
This set is naturally identified to $\setZ_2$: any sequence 
$\left(z_n\right )_{n\in\setN}\in\partial T$ is uniquely 
associated to a sequence $(a_n)_{n\geq 0}$ with $a_n\in\{0,1\}$ such that
$$
z_n=\sum_{m=0}^{n-1}a_m2^{m},
$$
and therefore to $q= \dots a_n \dots a_1 a_0\in \setZ_2$. 


Note that,  for two ends $x$ and $x'$, the  2-adic distance measures their  proximity with regards to the tree, more precisely

$$
n = - \log_2 \abs {x'-x } _2
$$
 is the index of the generation at which the corresponding  paths splitted.

%

As $V$, the set of vertices of $T$,  identifies with the disjoint union of the $(\setZ/ 2^n\setZ)$'s, any pressure field $p \in \setR^V$ can be seen as a sequence  $(p_n)_{n\in \setN}$, with 
 $p_n \in \setR^{\setZ/2^n\setZ }$.
We define a ``piecewise constant'' function $\tilde p_n$ over $\setZ_2$ as
\begin{equation}
\label{eq:deftp}
\tilde p_n (x) = p_n(a) \quad \forall x \in a  +  2^{n}\setZ_2\virg a  \in \setZ/2^n\setZ.
\end{equation}
  As an example, consider a field $p$ which takes values $-1$ and $1$ at $x_1^0$ and $x_1^1$, respectively. 
 Then $\tilde p_1$ is $-1$ on $2\setZ_2$ and $1$ on $1+2\setZ_2$.   
%
Note that $\tilde p_n$ lies in the Schwartz space $\S$ (see Definition ~\ref{def:S}
in the appendix).
The trace of $p$ on $\partial T = \setZ_2$ will be defined as the limit of $\tilde p_n$ as $n$ go to infinity, whenever it exists in some sense.


\subsection*{Trace operator onto $L^2$}
We consider from now on a  dyadic resistive tree $T_r = (V,E,r)$, where $r$ is the collection of edge resistances $(r(e))_{e\in E}$.
$$
r = \left ( r(e_n^k) = r_n^k\virg n \in \setN \virg
 0 \leq k \leq 2^n-1\right ) .
$$

Let us denote by $H^1(T_r)$ the set of all those functions with finite Dirichlet energy
$$
H^1(T) = \left \{ p \in \setR^V\virg \abs { p } _{H^1(T)} 
 < +\infty
\right \}.
$$
with the semi-norm
$$
\abs { p }   _{H^1} ^2 := 
\sum_{(x,y)=e\in E} \frac{ \abs {p(y)-p(x) } ^2  } {r(e)}
= \sum_{n=1}^\infty  \sum_{k=0}^{2^n-1}  \frac{ \abs {p(x_n^k)-p\left ( \varphi_{n-1}^n(x_n^k)\right )  } ^2  } { r_n^k}.
$$
Note that in the case $r(e)$ is constant in each generation ($r_n$ at generation $n$), the $H^1$ semi-norm reads
$$
\abs { p } _{H^1} ^2 = \sum_{n\geq 1} \frac {2^n}{r_n} \left\| \tilde p_{n} - \tilde p_{n-1} \right\|_{L^2(\ZZ)}^2,
$$
where $\tilde p_n$ is defined by~(\ref{eq:deftp}) and where $L^2(\ZZ)$ is defined with respect the Haar measure $\mu$ , see Definition \ref{def:haar} in the appendix.
It is a direct consequence of the fact that $\mu\left (  2^n \setZ_2\right ) = 2^{-n}$.

The following proposition allows to define a trace of $H^1$ functions over $T$ as soon as some condition on the resistances is met~:
\begin{proposition}
\label{prop:traceL2}
Let $T_r$ be a resistive dyadic tree, with $r = (r_n^k)$.
Assume
\begin{equation}
\label{eq:condrL2}
\sum_{n\geq 0}  \frac {1 }{2^n} \max_{k} r_n^k < +\infty.
\end{equation}
Then $(\tilde p_n)$ (defined by~(\ref{eq:deftp})) converges strongly to some $\tilde p \in L^2(\setZ_2)$.
The linear operator 
$\gamma_0 \; : \; p\longmapsto \tilde p$
is in  $\mathcal {L} \left (H^1(T) , L^2(\setZ_2)\right ) $.
\end{proposition}

\dem
It suffices us to show that 
\begin{equation} \label{cont:trace} \sum_{n=1}^\infty \left\| \tilde p_n-\tilde p_{n-1} \right\|_{L^2(\setZ_2)} \lesssim \|p\|_{H^1(T)}.\end{equation}
Using Cauchy-Schwarz inequality with coefficients $\beta_n$ (to be  chosen later), we have~:
\begin{align}
\sum_{n=1}^\infty \left\| \tilde p_n- \tilde p_{n-1} \right\|_{L^2(\setZ_2)} & \leq \left( \sum_n \beta_n \left\| \tilde p_n- \tilde p_{n-1} \right\|_{L^2(\setZ_2)}^2 \right)^{1/2} \left( \sum_n \beta_n^{-1}\right)^{1/2} \nonumber \\
 & \leq \left( \sum_n \beta_n \left\| \tilde p_n- \tilde p_{n-1} \right\|_{L^2(\setZ_2)}^{2} \right)^{1/2} \left( \sum_n \beta_n^{-1}\right)^{1/2}. \label{cauchy}
\end{align}
For the first term, it comes from the definition
$$ \left\| \tilde p_n- \tilde p_{n-1} \right\|_{L^2(\setZ_2)}^2 =
 \sum_{i=0 }^{2^n-1} \left|p(x_n^i)-p\left ( \varphi_{n-1}^n(x_n^i)\right )  \right|^2  2^{-n}.$$
Hence,
\begin{align*}
\lefteqn{\left( \sum_n \beta_n \left\| \tilde p_n-\tilde p_{n-1} \right\|_{L^2(\setZ_2)}^{2} \right)^{1/2} \leq} & & \\
 & & \left( \sum_n \beta_n 2^{-n} \sum_{i=0 }^{2^n-1} \left| p(x_n^i)-p\left (  \varphi_{n-1}^n(x_n^i)\right )  \right|^2 \right)^{1/2}.
\end{align*}
We choose the coefficients $\beta_n$ such that this last quantity is bounded by $\|p\|_{H^1(T)}$, i.e. such that
$$ \beta_{n} 2^{-n} = \inf_{0\leq i \leq 2^n-1} \frac{1}{r_{n}^i} = \frac{1}{\max_i r_{n}^i}.$$
With this particular choice, the first term in (\ref{cauchy}) is bounded by $\|p\|_{H^1(T)}$. Moreover the second term in (\ref{cauchy}) is finite thanks to~(\ref{eq:condrL2}) and
$$ \beta_n = \frac{2^n}{ \max_i r_{n}^{i}},$$
from which we deduce Inequality~(\ref{cont:trace}) and  the proposition. 
\findem

\begin{remark} 
 We have a similar result for the $L^p(\setZ_2)$ space instead of $L^2(\setZ_2)$.
For $p\geq 2$, if
\begin{equation} \sum_{n\geq 0}  \frac {1 }{2^n} \left[\max_{k} r_n^k\right]^{p/2} < +\infty \label{eq:condrLp} \end{equation}
then $(\tilde p_n)_n$ converges strongly in $L^p(\setZ_2)$. 
 Operator 
$\gamma_0 $ belongs to  $\mathcal {L} \left (H^1(T) , L^p(\setZ_2)\right ) $. 
\end{remark}
\begin{remark}
Condition~(\ref{eq:condrLp}) is met as soon as $ \max_{k} r_n^k \leq \alpha^n$ with $\alpha < 4^{1/p}$.
As for the infinite version of the actual human lungs, resistances vary like $\alpha^n$ with $\alpha \approx 1.6$, so that such a  trace operator can be defined properly in $L^p(\setZ_2)$ for $p\leq 2.9$.
\end{remark}

\section{Sobolev spaces in $\ZZ$ and precised trace theorems}
\label{sec:sob}

As in the case of Sobolev spaces over domains in $\setR^d$, it is natural to expect some regularity of functions in $\gamma_0(H^1(T))$.
It will be expressed in terms of $H^s$ regularity. As the functions we consider here are restricted to $\setZ_2$, standard Fourier transform
$$
\hat f (\xi) = \int _{\setQ_2} e^{-2i\pi x\xi } f(x) \, d\mu.
$$
does not depend on the integer part of $\xi$ (i.e. $\hat f (\xi) = \hat f (\xi')$ as soon as $\xi'-\xi \in \setZ_2$). As a consequence, the appropriate notion is a reduced Fourier transform (in the same way Fourier transform is replaced by Fourier series for periodic functions over $\setR$). This reduced Fourier transform, or Fourier series,  is defined as 
$$
\F(f) (\lambda)= \int_{\setZ_2}  e^{-2i\pi \lambda x }  f(x) d\mu(x),
$$
where $\lambda $ runs over $\Lambda = \setQ_2 / \setZ_2$ which identifies to $\setZ[1/2]$. Sobolev space of index $s$ is then defined as the set of all those functions such that 
$$
\| f \| _{H^s(\setZ_2)} := \left (
\sum_{\lambda \in \Lambda} \left ( 1 + \abs {\lambda}   _2  \right ) ^{2s} \abs {\F(f)(\lambda)}^2 
\right ) ^{1/2}
$$
is finite. Note that both notions are consistent : given $f \in L^2(\setZ_2)$, if  we define $\hat f$ as the standard Fourier transform of 
the extension of $f$ by $0$ on $\setQ_2 \setminus \setZ_2$, we have
$$
\hat f (\xi) = \F(f)(\lambda)
$$
as soon as $\xi - \lambda \in \setZ_2$. We refer the reader to Appendix~\ref{sec:fourier} for  more details on the underlying framework.
 
 \begin{proposition}\label{proptrhs}
Let $T_r$ be a resistive dyadic tree, with $r = (r_n^k)$.
Assume
\begin{equation}
\label{eq:condrHs}
\sum_{n\geq 0}  \frac {\max_{k} r_n^k  }{2^{n(1-2s)}} < +\infty, 
\end{equation}
for some $s >0$.
Then   $\gamma_0$ (defined by Prop.~\ref{prop:traceL2}) maps continuously  $H^1(T_r)$ onto $H^s(\setZ_2)$.
\end{proposition}

\dem  We follow the same ideas as for Proposition \ref{prop:traceL2}, proving
\begin{equation} \label{cont:trace2} \sum_{n=1}^\infty \left\| \tilde p_n-\tilde p_{n-1} \right\|_{H^s(\ZZ)} \lesssim \|p\|_{H^1(T)}.\end{equation}
In order to check this claim, we have to estimate the Sobolev norm of
$$
 \tilde p_n-\tilde p_{n-1} = \sum_{j\in \Z_2/2^{n}\Z_2} \left[p(x_n^j)-p\left (\varphi_{n-1}^n(x_n^j)\right ) \right]{\bf 1}_{j+ 2^{n}\ZZ}.
 $$

We fix $n$ and write $\phi_n^j:={\bf 1}_{j+ 2^{n}\ZZ}$. 
Using $\widehat{{\bf 1}_{\Z_2}}={\bf 1}_{\Z_2}$ (see Prop.~\ref{prop:ex} in Appendix) and a change of variable, we get
$$ \widehat{\phi_n^j}(\xi) = 2^{-n}e^{-2i\pi j\xi} {\bf 1}_{\ZZ}(2^{n}\xi).$$
Hence for $j\neq j'$, we obtain
\begin{align*}
 (\phi_n^j,\phi_n^{j'})_{H^s} & =
 4^{-n}\int_{\QQ} \left(1+|\xi|_2^2\right)^{s} 
e^{-2i\pi (j-j')\xi} {\bf 1}_{\Z_2}(2^{n}\xi)d\mu(\xi)\\
& = 
  4^{-n} \int_{|\xi |_2\leq 2^{n} } \left(1+|\xi |_2^2\right)^{s} 
e^{-2i\pi (j-j')\xi} d\mu(\xi) \\
 & =  4^{-n} \sum_{k=-\infty}^{ n} \left(1+2^{2k}\right)^{s} \int_{|\xi|_2=2^{k}} 
e^{-2i\pi (j-j') \xi} d\mu(\xi) \\
 & =  4^{-n} \sum_{k=-\infty}^{ n} \left(1+2^{2k}\right)^{s} \int_{|\eta|_2=2^{k}|j-j'|_2} 
e^{-2i\pi\eta} \frac{d\mu(\eta)}{|j-j'|_2} \\
 & \lesssim  4^{-n} \sum_{2^k |j-j'|_2 \leq 1 } \left(1+2^{2k}\right)^{s} 2^k \\
 & \lesssim \frac{4^{-n}}{ |j-j'|_2^{1+2s}}.
\end{align*}
Moreover for $j=j'$, we directly have
\begin{align*}
 (\phi_n^j,\phi_n^{j'})_{H^s} & = 4^{-n} \int_{|\xi |_2\leq 2^{n} } \left(1+|\xi |_2^2\right)^{s} d\mu(\xi) \\
 & =  4^{-n} \sum_{k=-\infty}^{n} \left(1+2^{2k}\right)^{s} 2^{k} \\
 & \lesssim  4^{-n} \left(1+2^{2n}\right)^{s} 2^n \\
 & \lesssim 2^{n(-1+2s)}.
\end{align*}
Finally, it comes
\begin{align*}
& \| \tilde p_n- \tilde p_{n-1}\|_{H^s(\Z_2)}^2    \\
\leq & \sum_{j \neq j'} \left[p(x_n^j)-p\left ( \varphi_{n-1}^n(x_n^j)\right )\right]\left[p(x_n^{j'})-p\left (\varphi_{n-1}^n(x_n^{j'})\right )\right] \frac{4^{-n}}{ |j-j'|_2^{1+2s}} \\
 & + \sum_{j} \left[p(x_n^j)-p\left (\varphi_{n-1}^n(x_n^j)\right )\right]^2 2^{n(-1+2s)}.
\end{align*}
Splitting the first sum with $xy\leq x^2 + y^2$, it comes two symmetrical sums. Then using
$$ \sum_{k=1}^{2^n-1} \frac{4^{-n}}{ |k|_2^{1+2s}} = 4^{-n} \left( 2^{n-1} + 2^{n-1+2s} + \cdots + 2^{n-1+2sn}\right) \simeq 2^{n(-1+2s)}$$
we prove that
$$ \| \tilde p_n- \tilde p_{n-1}\|_{H^s(\Z_2)}^2 \leq \sum_{j} \left[p(x_n^j)-p\left (\varphi_{n-1}^n(x_n^j)\right )\right]^ 2 2^{n(-1+2s)}. $$
Then we conclude the proof by the same way as for Proposition \ref{prop:traceL2} .\findem

\medskip \noindent Let us treat the particular case of a finite resistance sub-geometrical tree: the resistances satisfy $r_{n}^i \leq \alpha^n$ for some parameter
 $\alpha\in(1,2)$. Assumption (\ref{eq:condrHs}) is satisfied if and only if we are in the 
 sub-critical case:
\begin{equation} \label{eq:scritique} s < s_\alpha:=(1- \log_2(\alpha))/2. \end{equation}
For $s\in[0,s_\alpha)$, we can obtain a bound for the convergence velocity of $\tilde p_n$ to $\gamma_0(p)$. More precisely,  
$$ \| \tilde p_n-\gamma_0(p) \|_{H^s(\ZZ)} \lesssim \left(\frac{\alpha}{2^{1-2s}}\right)^n \| f\|_{H^1(T)}=4^{n(s-s_\alpha)} \| f\|_{H^1(T)}.$$
The following Theorem gives a positive result in the critical case $s=s_\alpha$ for the sub-geometrical trees (we have a positive result of convergence without a precise estimate of the convergence velocity)

\begin{theorem} Let us consider a sub-geometrical tree $r_{n}^{i}\leq \alpha^n$ with $\alpha\in (1,2)$. Then  $\gamma_0$
 (defined by Prop.~\ref{prop:traceL2}) maps continuously  $H^1(T_r)$ onto $H^{s_\alpha}(\setZ_2)$, with $s_\alpha = (1-\log_2 \alpha) / 2$.
\end{theorem}

\dem We claim first that for every function $p\in H^1(T_r)$, the trace $\gamma_0(p)$ belongs to the limit space $H^{s_\alpha}(\ZZ)$
 and let us first conclude. From that, the linear operator $\gamma_0$ is acting from $H^1(T_r)$ to $H^{s_\alpha}(\ZZ)$. 
Since the previous continuity from $H^1(T_r)$ in $L^2(\ZZ)$, it is easy to check that the graph of $\gamma_0$ is closed 
in $H^1(T_r)\times H^{s_\alpha}(\ZZ)$. Then Banach's Theorem of closed graph implies the desired continuity  of  $\gamma_0$. \\
It remains  to prove that $\gamma_0$ maps $H^1$ onto  $H^{s_\alpha}(\ZZ)$. Let us fix a function $p\in H^1(T_r)$.
We are going to show that the sequence $(\tilde p_n)_n$ is  Cauchy in $H^{s_\alpha}(\ZZ)$.
Let $n<m$ be two integers and write 

$$ \tilde p_n(x) := \sum_{i=0}^{2^n - 1} p(x_n^i){\bf 1}_{i+ 2^n\Z_2}(x) =
  \sum_{i=0}^{2^n- 1} p(x_n^i) \ \sum_{\genfrac{}{}{0pt}{}{0\leq j < 2^m}{j\equiv i \; [2^n]}} {\bf 1}_{j+ 2^m\Z_2}(x).$$

  \begin{figure}
\psfragscanon

\psfrag{n}{$n$}
\psfrag{m}{$m$}
\psfrag{xni}{$x_n^i$}
\psfrag{xnip}{$x_n^{i'}$}
\psfrag{xmj}{$x_m^{j}$}
\psfrag{xmjp}{$x_m^{j'}$}
\psfrag{jin}{$j\equiv i \; [2^n]$}
\psfrag{jpipn}{$j'\equiv i' \; [2^n]$}
\centerline{
\includegraphics[width=1\textwidth]{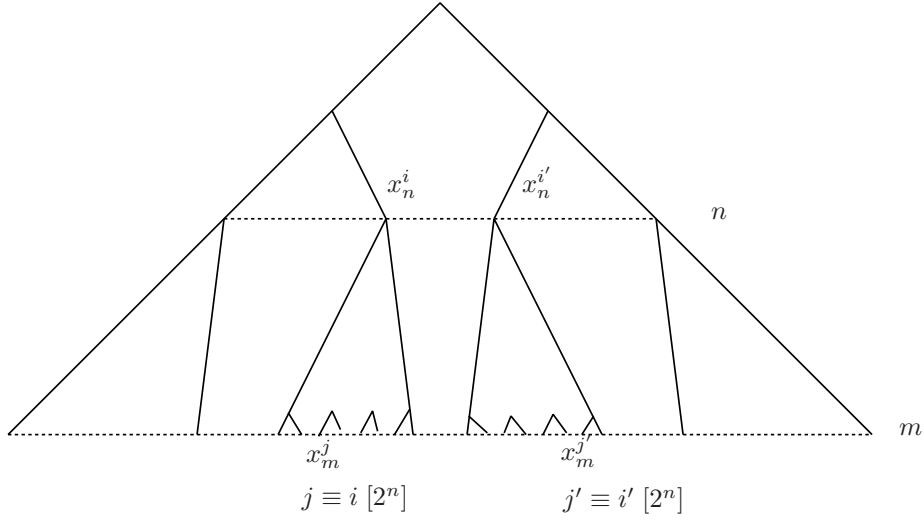}
}
\label{fig:ij}
\caption{Notations}
\end{figure}
  
Hence
$$ \tilde p_n(x)- \tilde p_m(x) =
 \sum_{i=0}^{2^n - 1} 
  \sum_{\genfrac{}{}{0pt}{}{0\leq j < 2^m}{j\equiv i \; [2^n]}} 
  \left[ p(x_n^i) -p(x_m^j) \right] {\bf 1}_{j+ 2^m\Z_2}(x).$$
As previously (see Fig.~\ref{fig:ij} for the meaning of notations), we get
\begin{align*}
 \| \tilde p_n-  \tilde p_m \|_{H^{s_\alpha}(\ZZ)}^2 & \\
& \hspace{-2cm} = 
 \sum_{i=0}^{2^n - 1} 
  \sum_{\genfrac{}{}{0pt}{}{0\leq j < 2^m}{j\equiv i \; [2^n]}} 
 \sum_{i'=0}^{2^n - 1} 
  \sum_{\genfrac{}{}{0pt}{}{0\leq j'<2^m}{j'\equiv i' \; [2^n]}} 
\ \left[ p(x_n^i)-p(x_m^j) \right]^ 2 \frac{4^{-m}}{ |j-j'|_2^{1+2s_\alpha}} \\
& \hspace{-2cm} \lesssim \sum_{i=0}^{2^n - 1}  \ \sum_{\genfrac{}{}{0pt}{}{0\leq j < 2^m}{j\equiv i \; [2^n]}} 
\left[ p(x_n^i)-p(x_m^j) \right]^ 2 2^{m(-1+2s_\alpha)}.
\end{align*}
To control $ p(x_n^i)-p(x_m^j) = p(\varphi_n^m(x_m^j))-p(x_m^j) $, we use triangle inequality on the telescopic series along the path between $x_n^i$ and $x_m^j$, as follows
\begin{align*}
\left[  p(x_n^i)-p(x_m^j) \right]^ 2 & = \left[ \sum_{k=n}^{m-1} p(\varphi_k^m(x_m^j))-p(\varphi_{k+1}^m(x_m^j)) \right]^ 2 \\
 & \lesssim \left[ \sum_{k=n}^{m-1} \frac{\left| p(\varphi_k^m(x_m^j))-p(\varphi_{k+1}^m(x_m^j)) \right|^2}{r_{k+1}^{j}} \right] \alpha^m.
\end{align*}
We have used Cauchy-Schwarz inequality at the last step, the assumption of sub-geometric increasing of the resistances, the fact that $\alpha>1$. Note that $r_{k+1}^j$, which is a priori not defined as $j$ might be larger than $ 2^{k+1}$, represents obviously $r_{k+1}^\ell$, where $0\leq \ell  < 2^{k+1}$, with $\ell \equiv j \; [2^{k+1}]$.
By definition of $s_\alpha$, we deduce 
\begin{align*}
 \| \tilde p_n- \tilde p_m \|_{H^{s_\alpha}(\ZZ)}^2 & \\
 & \hspace{-2cm} \lesssim  \sum_{i=0}^{2^n - 1}  \   \sum_{\genfrac{}{}{0pt}{}{0\leq j< 2^m}{j\equiv i \; [2^n]}} 
 \left[ \sum_{k=n}^{m-1} \ \frac{ \left|p(\varphi_k^m(x_m^j))-p(\varphi_{k+1}^m(x_m^j)) \right| ^2}{r_{k+1}^{j}} \right] \alpha^m 2^{m(-1+2s_\alpha)} \\
 & \hspace{-2cm} \lesssim  \sum_{i=0}^{2^n - 1}  \   \sum_{\genfrac{}{}{0pt}{}{0\leq j<  2^m}{j\equiv i \; [2^n]}}  \frac{\left|p(\varphi_k^m(x_m^j))-p(\varphi_{k+1}^m(x_m^j)) \right|^2}{r_{k+1}^{j}} \\
 & \hspace{-2cm} \lesssim \| p \|_{H^1(T_n)}^2,
\end{align*}
where $T_n$ is the subnetwork corresponding to the  the set of generations $k$ with $k\geq n$. As $ \| p \|_{H^1(T_n)}$ goes to $0$, we have also proved that $(\tilde p_n)_n$ is a Cauchy sequence in $H^{s_\alpha}(\ZZ)$ so that $\gamma_0(p)$ belongs to this space. \findem

\section{DN and ND operators}
\label{sec:DN}
Given a pressure field on $\ZZ$ (seen as the set of ends of the tree $T_r$), we are  interested in the fluxes which it drives through $T_r$ (Dirichlet problem), and in particular in the quantity of air which exits the tree through its boundary, which amounts to solve the  Dirichlet to  Neuman (DN) problem associated with the tree. 
 The reciprocal (ND) mapping is straightforward to obtain, as detailed in~\cite{GMM}.  Indeed as soon as the global flux is known, it identifies with the flux through the first edge, which gives the pressure at generation $0$ (pressure at the root is $0$). All pressures can be  computed recursively in a similar manner, for the flux through any subtree is known.
Following this procedure in the case of  a regular tree (resistance $r_n$ at generation $n$), given a flux field $u \in \setR ^{\ZZ}$ (the regularity of which will be addressed  later),
 pressure at end $a\in \ZZ$ writes formally
$$
p(a) = \sum_{n=0}^{+\infty} R_n \int_{\abs { x-a} _2 = 1/2^n} u(x) \, d\mu(x),
$$
where $R_n = r_0+r_1+\dots+r_n$ is the cumulated resistance.
If one assumes  geometric growth of the resistances according to some $\alpha \in (1,2)$, more precisely
$$
r_0 = 1\virg r_n = \alpha ^{n-1} (\alpha-1),
$$
one obtains $R_n = \alpha^n$, and consequently 
$$
p(a) = \int_{\ZZ} \frac { u(x)\phantom{\log_2 \alpha}}{\abs{x-a} _2 ^{\log_2 \alpha} } \,d\mu(x)
 =  \zeta(\beta) \ \tilde k_\beta \star u(a),
$$
where $\tilde k_\beta$ is  the Riesz kernel on $\ZZ$ (see Section~\ref{sec:riesz} for a brief presentation of these multipliers) 
$$
\tilde k^\beta(x) := \frac {2} {\zeta(\beta)} \abs { x } _2^{\beta -1}\virg \hbox {with }\beta := 1-\log_2 \alpha>0, 
$$ 
and $\zeta(\beta) = (1 - 2^{-\beta})^{-1}$ is the local zeta function.

 \begin{proposition} (Neuman-Dirichlet operator)\\
 The Riesz operator $\widetilde{\mathcal R}^\beta$ (corresponding to the convolution by $\tilde k^\beta$ on $\ZZ$) maps continuously $H^{-s}(\ZZ)$ onto $H^s(\ZZ)$, with 
 $s = \beta/2$.
 \end{proposition}

From a functional point of view, we detail in Section \ref{sec:riesz} that the Riesz operators $\widetilde{\mathcal R}^\beta$ corresponds to some power of a Laplacian operator $\widetilde{\mathcal R}^\beta = \Delta^{-\beta/2}$. So it is natural to expect that it maps continuously $H^{-s}(\ZZ)$ onto $H^{-s + \beta }(\ZZ)$ for all exponent $s>0$ (the desired result is a particular case of this property). Let us give a more detailed proof.

\dem We refer the reader to Section \ref{sec:sobolevZZ} for a presentation of Sobolev spaces on $\ZZ$. 
Thanks to~(\ref{eq:sobZZ}) we have
$$ \|\widetilde{\mathcal R}^\beta(f)\|_{H^s(\ZZ)} \simeq \left(\sum_{\lambda\in \Lambda} \left ( 1 + \abs { \lambda} _2 \right ) ^{2s} | \F(\tilde k^\beta)(\lambda) \F(f) (\lambda) | ^2 \right)^{1/2},$$
with by convention $\abs { \lambda} _2=1$ for $\lambda=0\in \Lambda$.
Moreover, (\ref{rel:fourier2}) gives
$$ \F(\tilde k^\beta)(\lambda)=|\lambda|_2^{-\beta}. $$
So it comes
$$ \|\widetilde{\mathcal R}^\beta(f)\|_{H^s(\ZZ)} \simeq \left(\sum_{\lambda\in \Lambda} \left ( 1 + \abs { \lambda} _2 \right ) ^{2s} |\lambda|_2^{-2\beta} |\F(f) (\lambda) | ^2 \right)^{1/2}.$$
For $\lambda=0$, by convention $\abs { \lambda} _2=1$ so 
$$\left ( 1 + \abs { \lambda} _2 \right ) ^{2s} |\lambda|_2^{-2\beta} \simeq \left ( 1 + \abs { \lambda} _2 \right ) ^{2(s-\beta)}.$$
For $\lambda\neq 0$ then  $\abs { \lambda} _2>1$ so
$$\left ( 1 + \abs { \lambda} _2 \right ) ^{2s} |\lambda|_2^{-2\beta} \simeq \left ( 1 + \abs { \lambda} _2 \right ) ^{2(s-\beta)}.$$
We also conclude that as expected
$$ \|\widetilde{\mathcal R}^\beta(f)\|_{H^s(\ZZ)} \simeq \| f\|_{H^{s-\beta}(\ZZ)},$$
which for $s=\beta/2$ gives us the desired estimate.
\findem

\medskip
\noindent According to Corollary \ref{cor:Riesz}, we know that the Riesz multiplier $\widetilde{\mathcal R}^\beta$ is invertible in distributional sense or in $L^2$ sense and $(\widetilde{\mathcal R}^\beta)^{-1} = \widetilde{\mathcal R}^{-\beta}$. 
It gives an explicit expression of the Dirichlet-Neuman operator  for the regular tree, i.e. the mapping
$$
\hbox{Pressure field }\longmapsto \hbox{ Fluxes}, 
$$
which is the core of the ventilation process.

 
 \section{Embedding onto a domain of $\setR^d$}
 \label{sec:embed}
 One expected outcome of this approach is to provide a sound functional framework for the coupling of a resistive tree with an elastic medium onto which it is embedded (see~\cite{GMM} for a first coupled model in one-dimension, and~\cite{baf}  for an homogenized description of a foamy like medium) . We must say that the 2-adic viewpoint does not allow to improve significantly the results which are presented in~\cite{msv} on this matter. Considering a mapping from $\ZZ$ onto a  domain in $\setR^d$ (which models the way our infinite tree is imbedded on the physical space, i.e. the actual lung),  we simply give here a property which allows to describe how Sobolev regularity of a function on  the  domain can be transported  back to its  $\ZZ$ counterpart, as soon as   some H\"older regularity of the mapping is verified. 
  
Let $\Omega$ be an open set of $\setR^d$ (equipped with its Euclidean structure and Lebesgue measure) with $\abs {\Omega }  = 1$,  and $\phi$ a measure-preserving mapping from $\ZZ$ onto $\Omega$~: for every measurable set $A\subset \Omega$
$$ |\mu(\phi^{-1}(A))| = \abs A .$$
We then define the following operator~:
$$ T_\phi(f):= f\circ \phi.$$

\begin{proposition}\label{iso}
 For all exponent $p\in[1,\infty]$, $T_\phi$ continuously acts from $L^p(\Omega)$ to $L^p(\Z_2)$. More precisely $T_\phi$ is an isometry~:
$$\|T_\phi(f)\|_{L^p(\ZZ)} = \|f\|_{L^p(\Omega)}.$$ 
 \end{proposition}

\dem Assume first that $p<\infty$. Using 
$ \mu\left(\left\{x, \ |f\circ \phi(x)|> t\right\} \right)  =  \left|\left\{y,\ |f(y)|> t\right\}\right| $, we obtain  immediately
$$
 \|T_\phi(f)\|_{L^p(\ZZ)}^p = \|f\|_{L^p(\Omega)}^p. 
 $$
The $p=\infty$ case follows by having $p$ go to infinity.  
%
%
\findem

\noindent We are now looking for condition on $\phi$ such that $T_\phi$ keeps some regularity. 

\begin{proposition}\label{injreg}
 If $\phi:\ZZ \to \Omega$ is  $1/d$-H\"olderian, then $T_\phi$ is continuous from $H^{ds}(\Omega)$  to $H^s(\Z_2)$ for all $s\geq 0$. 
 \end{proposition}

\dem We use the characterization of Sobolev spaces, given by Proposition \ref{prop:sob}
$$ \|T_\phi(f) \|_{H^s(\ZZ)} \lesssim \|T_\phi(f)\|_{L^2(\ZZ)} + \left( \int_{\ZZ} \int_{\ZZ} \frac{|f\circ \phi (x)-f \circ \phi (y)|^2}{|x-y|_2^{1+2s}} d\mu(x)d\mu(y) \right)^{1/2}.$$
The previous proposition yields
$$ \|T_\phi(f)\|_{L^2(\ZZ)} = \|f\|_{L^2(\Omega)}.$$
The $1/d$-H\"olderian regularity of $\phi$ gives
$$ |\phi(x)-\phi(y)| \lesssim |x-y|_2^{1/d},$$
which implies
\begin{align*}
 \left( \int_{\ZZ} \int_{\ZZ} \frac{|f\circ \phi (x)-f \circ \phi (y)|^2}{|x-y|_2^{1+2s}} d\mu(x)d\mu(y) \right)^{1/2}& \lesssim&
  \\
 \left( \int_{\ZZ} \int_{\ZZ} \frac{|f\circ \phi (x)-f \circ \phi (y)|^2}{|\phi(x)-\phi(y)|^{d+2ds}} d\mu(x)d\mu(y) \right)^{1/2}.&&
\end{align*}
Then, we use again the previous proposition, to deduce
$$ \left( \int_{\ZZ} \int_{\ZZ} \frac{|f\circ \phi (x)-f \circ \phi (y)|^2}{|x-y|_2^{1+2s}} d\mu(x)d\mu(y) \right)^{1/2} \lesssim 
\left( \int_{\Omega} \int_{\Omega} \frac{|f(x)-f(y)|^2}{|x-y|^{d+2ds}} dxdy \right)^{1/2}.$$
Using the well-known characterization of local Sobolev spaces, we get the desired estimate
$$ 
\|T_\phi(f) \|_{H^s(\ZZ)} \lesssim \|f\|_{L^2(\Omega)} + \left( \int_{\Omega} \int_{\Omega} \frac{|f(x)-f(y)|^2}{|x-y|^{d+2ds}} dxdy \right)^{1/2} \lesssim \|f\|_{H^{ds}(\Omega)}.$$
\findem
 

 \appendix


 \section{Fourier transform on $\QQ$ and Fourier series on $\ZZ$}
  \label{sec:fourier}
 
 \begin{definition} 
 \label{def:haar} The set $\QQ$ endowed with its $2$-adic distance $d(x,y)=|x-y|_2$ is a locally compact group. It  owns a Haar measure  $\mu$ 
which  satisfies for every $x\in \QQ$ and $k\in \Z$
 $$ \mu\left( x + 2^k \ZZ\right) = 2^{-k} \mu\left(\ZZ\right) = 2^{-k}.$$
 This measure defines  a probability measure on $\ZZ$.
  \end{definition}
 
\begin{definition}[Schwartz space]
 \label{def:S}
 The Schwartz space  $\S(\QQ)$ is defined as the space of all those functions  which are compactly supported and locally constant.
 It is spanned by characteristics functions of balls $\one_{a + 2^m \ZZ}$, $a\in  \setQ_2$, $m\in \setZ$.
 \end{definition}

\begin{definition}
  \label{def:ft}
For any  $f \in \S(\QQ)$, its Fourier transform is defined by 
$$
\xi \longmapsto   \hat f (\xi)  := \int_\QQ  e^{-2i\pi x \xi } f(x) \, d\mu(x).
$$
 \end{definition}
 Characteristic functions of closed balls containing (i.e. centered at) $0$ play the role of central Gaussian distributions in $\setR^d$~:

\begin{proposition} \label{prop:ex} We have
 $$
 \widehat{ {\bf 1}_{2^k\ZZ} } = 2^{-k}{\bf 1}_{2^{-k}\ZZ}.
  $$
  In particular $\widehat{ {\bf 1}_{\ZZ} } = {\bf 1}_{\ZZ} $.
\end{proposition}


\begin{proposition} \label{prop:formule} We have
$$ \int_{|x|\leq 2^k} e^{2i\pi x} d\mu(x) = 2^k{\bf 1}_{k\leq 0}$$
and
\begin{equation}
\label{eq:intspheres}
 \int_{|x|= 2^k} e^{2i\pi x} d\mu(x) = 2^{k-1}{\bf 1}_{k\leq 0}+ (-1){\bf 1}_{k=1}.
 \end{equation}
\end{proposition}

\noindent The Fourier transform on $\QQ$ satisfies to the same properties than the ones on ${\mathbb R}^d$ : 

\begin{theorem} \label{thm:plancherel}
The Fourier transform is an isometry on $L^2(\QQ)$~: for all $f\in\S(\QQ)$
$$ \| \hat f \|_{L^2(\QQ)} = \|f\|_{L^2(\QQ)}.$$
Consequently, Fourier transform can be extended by density to a continuous operator over $L^2(\QQ)$. 
\end{theorem}
 
 \begin{theorem} \label{thm:inversion}
The Fourier transform is invertible on $L^2(\QQ)$ and,  for any function $f\in L^2(\QQ)$, 
$$ \hat{ \hat f}(x) = f(-x).$$
\end{theorem}

\noindent We would like to finish this section by describing the theory of Fourier series. On the Euclidean space ${\mathbb R}$, it is more convenient to use Fourier series for functions supported on $[0,1]$, similarly we can define Fourier series for functions supported on $\ZZ$. This new operation is denoted by $\F$, we follow the same scheme as in the Euclidean framework with identifying ${\mathbb R}$ to $\QQ$ and $[0,1]$ to $\ZZ$. We shall  denote the countable set $\QQ/\ZZ$ by $\Lambda$. 

\medskip
\begin{definition} Let $\lambda\in \Lambda$ and $f\in \S(\ZZ)$ a function supported on $\ZZ$. We define
$$ \F(f)(\lambda) := \int_{\ZZ} f(x) e^{-2i\pi x \lambda} d\mu(x).$$
We note that $x$ belonging to $\ZZ$, the previous quantity is well-defined for $\lambda\in \Lambda:=\QQ/\ZZ$. 
\end{definition}

\noindent Then we have the following properties (coming from those of the whole Fourier transform)~:

\begin{proposition}
For $f\in \S(\ZZ)$, we have
$$ f(x) = \sum_{\lambda \in \Lambda} \F(f)(\lambda)e^{2i\pi x \lambda}$$
and
$$ \|f\|_{L^2(\ZZ)} = \sum_{\lambda \in \Lambda} |\F(f)(\lambda)|^2.$$
\end{proposition}
 
\section{Sobolev spaces and $A^{s}$ spaces over $\ZZ$.} \label{sec:sobolevZZ}
 
 Similarly to what is done on the Euclidean space, we define Sobolev spaces on $\ZZ$ and the equivalent of the $A^{s}$ approximate spaces used in article \cite{msv} (see~\cite{cohen} for the definition and main properties of the $A^s$ spaces in the euclidean case).
\begin{definition}[Sobolev spaces]
 \label{def:sobolev}
For any $s \in {\mathbb R}$, we define for a Schwartz function $f\in\S(\ZZ)$
$$ \|f\|_{H^s(\ZZ)} := \left(\int_{\setQ_2} \left ( 1 + \abs { \xi} _2 \right ) ^{2s} | \hat f (\xi) | ^2 \, d\mu(\xi) \right)^{1/2}.$$
Then $\|\cdot\|_{H^s(\ZZ)}$ is a norm and we define the Sobolev space $H^s(\ZZ)$ as the 
completion of $\S(\ZZ)$ for this norm.
Then for $s\geq 0$,  $H^{-s}(\ZZ)$ can be identified to the dual space $H^s(\ZZ)'$.
The Sobolev spaces are Hilbert spaces.
  \end{definition}

\noindent In the previous section, we have seen that for functions defined on $\ZZ$, we can use Fourier series instead of Fourier transform. For such functions, if $\xi\in \ZZ$ then $\widehat{f} (\xi) = \int f d\mu = \F(f)(0)$ else
 the quantity $\widehat{f} (\xi)$ depends only on the class of $\xi\neq 0 \in \Lambda$ and
$$ \int_{\ZZ} \left ( 1 + \abs { \xi + \omega } _2 \right ) ^{2s} \, d\mu(\omega) = \left ( 1 + \abs { \xi } _2 \right ) ^{2s}.$$
Consequently, we have the following representation of the Sobolev norms, using Fourier series. For a function $f$ supported on $\ZZ$
\begin{equation} \label{eq:sobZZ} \|f\|_{H^s(\ZZ)} \simeq \left(\sum_{\lambda\in \Lambda} \left ( 1 + \abs { \lambda} _2 \right ) ^{2s} | \F(f) (\lambda) | ^2 \right)^{1/2},\end{equation}
with by convention $\abs { \lambda} _2=1$ for $\lambda=0\in \Lambda$.

Regularity estimates in~\cite{msv} rely  on $A^s$ norms (see~\cite{cohen} for a full description of this general setting). Although it does not play a central role in the present approach, we present here an equivalent definition of Sobolev spaces based on this definition of regularity. 

\begin{definition}[$A^{s}$ spaces] 
Let $n \in \mathbb{N}$, $V_{n}:= span ({\bf 1}_{\mathbb{Z}_{2}},{\bf 1}_{k+ 2^{j}\mathbb{Z}_{2}} ) _{j \leq n}$ and  let  
 $P_{n}: L^{2}(\ZZ) \to V_{n}$  be the projector onto  $V_{n}$ for the scalar product of $L^{2}( \ZZ)$.
Let $s > 0$. We define the  $A^{s}(\ZZ)$ space by
$$A^{s}(\ZZ):= \{ f \in L^{2}(\ZZ) \ \hbox{such that} \ \sum_{n=0}^{+ \infty} \|f-P_{n} f\|_{L^{2}(\ZZ)}^{2} 2^{2ns} <+ \infty .\}$$
The norm associated to this space is given by
$$\|f\|_{A^{s}}:= \|P_{0}f\|_{L^{2}(\ZZ)}+   \Big(\sum_{n=0}^{+ \infty} \|f-P_{n} f\|_{L^{2}(\ZZ)}^{2} 2^{2ns} \Big)^{\frac{1}{2}}.$$
\end{definition}
The following proposition  establishes the  link  between $A^s$ regularity and Sobolev regularity (Fourier setting). 

\begin{proposition}\label{pr2}
Let  $s > 0$. Then the following identification holds 
$$H^{s}(\ZZ)=A^{s}(\ZZ).$$
\end{proposition}

\dem

\noindent Let  $s>0$. Using the Plancherel's formula (see Theorem \ref{thm:plancherel}), we obtain that for all $n \in \mathbb {N}$
$$\|P_{n}u-u\|_{L^{2}(\ZZ)}=   \| \mathcal{F}(u -P_{n}u)\|_{L^{2}(\mathbb{Q}_{2})}.$$
But,  if $u \in L^{2}(\mathbb{Z}_{2})$, then  for all $n \geq 0$ 
\begin{equation}\label{1}
\hbox{supp} \ \mathcal{F}(P_{n}u) \subset B_{2}(0,2^{n}) \hbox{ and } \hbox{supp} \ \mathcal{F}(Id-P_{n}u) \subset {}^{c}B_{2}(0,2^{n}).
\end{equation}
Indeed, the first part of  property (\ref{1}) is a direct consequence of the explicit formula of $\widehat{\phi_{j}}$ 
(see the proof of Proposition~\ref{proptrhs}). Let $n \in \mathbb{N}$ and $0\leq k \leq 2^{n}-1\}$. To obtain the second part  of property (\ref{1}), 
 it is enough to prove 
 that for all functions $u \in L^{2}(\mathbb{Z}_{2})$ such that  $ \hbox{supp} \ u  \subset k+2^{n} \mathbb{Z}_{2}$ with
$$\int_{k+2^{n}\mathbb{Z}_{2}} u(x)\,d\mu=0,$$
 we have
$$ \hbox{supp} \ \mathcal{F}(u)  \subset {}^{c}B_{2}(0,2^{n}).$$
Let $\xi \in B_{2}(0,2^{n})$. Then for all $x \in k+2^{n} \mathbb{Z}_{2}$
$$e^{2i \pi x \xi}= e^{2i \pi k \xi}$$
and so 
$$ \widehat{u}(\xi)= \int_{k+2^{n}\mathbb{Z}_{2}} e^{2i \pi x \xi} u(x)\,d\mu= e^{2i \pi k \xi} \int_{k+2^{n}\mathbb{Z}_{2}}  u(x)\,d\mu=0$$
which ends the proof of~(\ref{1}). Applying~(\ref{1}), we obtain 
$$  \|\mathcal{F}(u -P_{n}u)\|_{L^{2}(\mathbb{Q}_{2})}= \|\hat{u}\|_{L^{2}({}^{c}B_{2}(0,2^{n}))}.$$
 We deduce that  
$$ \|P_{n}u-u\|_{L^{2}(\ZZ)}^{2}=\|\hat{u}\|_{L^{2}({}^{c}B_{2}(0,2^{n}))}^{2} \simeq \sum_{k=n+1}^{+\infty} 2^{-2sk}
  \int_{|\xi|_{2} = 2^{k}} (1+ |\xi|)^{2s}|\hat{u}|^{2}(\xi) d\xi.$$
and so
$$ \sum_{n=0}^{+ \infty}2^{2ns}\|P_{n}u-u\|_{L^{2}(\ZZ)}^{2} \simeq  \sum_{n=0}^{+\infty} \sum_{k=1}^{+\infty} a_{n-k} b_{k}  $$
where
$$a_{j}= \mathbb{I}_{j \leq - 1} 2^{2js}, \ b_{j}= \mathbb{I}_{j \geq 1}  \int_{|\xi|_{2} = 2^{j}} (1+ |\xi|)^{2s}|\hat{u}|^{2}(\xi) d\xi  .$$
Making the change of variables  $\ell=n-k$, we obtain 
$$ \sum_{n=0}^{+ \infty}2^{2ns}\|P_{n}u-u\|_{L^{2}(\ZZ)}^{2} \simeq \sum_{k=1}^{+\infty} \sum_{l=-\infty}^{+\infty}  a_{l} b_{k}$$ 
which ends the proof of Proposition \ref{pr2}. 
\findem


 \noindent Without requiring Fourier transform or frequential decomposition, we have a more geometrical characterization of Sobolev spaces (with positive index)~:
 
 \begin{proposition} \label{prop:sob} For all $s>0$ and $f\in \S(\ZZ)$, 
$$ \|f\|_{H^s(\ZZ)} \simeq \|f\|_{L^2(\ZZ)} + \left(\int_{\ZZ} \int_{\ZZ} \frac{|f(x)-f(y)|^2}{|x-y|_2^{1+2s}} d\mu(x)d\mu(y) \right)^{1/2}.$$
\end{proposition}

\dem Considering the double integral and properties of the Haar measure $\mu$, a change of variables yields 
$$\int_{\ZZ} \int_{\ZZ} \frac{|f(x)-f(y)|^2}{|x-y|_2^{1+2s}} d\mu(x)d\mu(y) = \int_{\ZZ} \frac{1}{|h|_2^{1+2s}} \int_{\ZZ} \left|f(x+h)-f(x)\right|^2 d\mu(x) d\mu(h).$$
Then according to Plancherel's inequality (Theorem \ref{thm:plancherel}), we get
$$ \int_{\ZZ} \left|f(x+h)-f(x)\right|^2 d\mu(x) = \int_{\QQ} |\hat f (\xi)|^2 |1-e^{2i\pi h\xi}|^2 d\mu(\xi).$$
Consequently
$$\int_{\ZZ} \int_{\ZZ} \frac{|f(x)-f(y)|^2}{|x-y|_2^{1+2s}} d\mu(x)d\mu(y) = \int_{\QQ} \theta(\xi) |\hat f (\xi)|^2 d\mu(\xi)$$
with
$$ \theta(\xi) := \int_{\ZZ} \frac{|1-e^{2i\pi h\xi}|^2}{|h|_2^{1+2s}} d\mu(h).$$
It also suffices to prove 
\begin{equation} \label{eq:equiv} (1+|\xi|_2^2)^{s} \simeq 1 + \theta(\xi). \end{equation}
First if $\xi\in \ZZ$, then $\theta(\xi)=0$ (as $h\xi$ would be an integer in the integral) ; (\ref{eq:equiv}) holds since $|\xi|_2\leq 1$. \\
Else for $\xi\notin \ZZ$, we denote by $p$ the negative integer satisfying $|\xi|_2=2^{-p}$, which is equivalent to $\xi \in 2^p(1+2\ZZ)$. For all $h\in\ZZ$ with $|h|_2\leq 2^{p}$, $h\in 2^{-p}(1+2\ZZ)$ and so $h\xi \in \Z_2$. Hence
\begin{align*}
 \theta(\xi)& =  \int_{|h|_2\geq 2^{p+1}} \frac{|1-e^{2i\pi h\xi}|^2}{|h|_2^{1+2s}} d\mu(h)
    \simeq  \sum_{k=p+1}^{0} 2^{-2ks} 
     \simeq 2^{-2ps} \simeq |\xi|_2^{2s},
 \end{align*}
 where we used that for a non positive integer $l$
$$ \int_{|x|_2= 2^l} e^{2i\pi x} d\mu(x) = 2^{l-1},$$
 due to Props.~\ref{prop:ex} and~\ref{prop:formule}. 
That concludes the proof of (\ref{eq:equiv}). \findem
 
 \section{Riesz kernels and associated multipliers} \label{sec:riesz}

We recall here some well-known properties concerning particular multipliers, namely Riesz kernels (see \cite{AKS, Haran, Haran2} for more details).

\begin{definition} For $\beta>0$, we consider the following function defined on $\QQ$
$$ k^\beta(\xi):=\frac{\zeta(1-\beta)}{\zeta(\beta)} |x|_2^{\beta-1}$$
and,  for $\beta<0$, 
$$ k^\beta(x):=p.v.\ \frac{\zeta(1+\beta)}{\zeta(-\beta)} |x|_2^{\beta-1},
$$
where $p.v.$  stands for the principal value based on the following representation formula:
$$ \int k^\beta \phi d\mu = \int k^\beta(x) \left[\phi(x)-\phi(0)\right] d\mu(x),$$
and  $\zeta$ is  the local zeta function~:
$$\zeta(\beta):= \zeta(1)\int_{\ZZ^*} |x|_2^\beta \frac{d\mu(x)}{|x|_2} = \frac{1}{1-2^{-\beta}}.$$
We denote by ${\mathcal R}^\beta$ the multiplier defined as the convolution operator by $k^\beta$.
\end{definition}

\noindent We have a precise description in the frequency space of the symbols associated to these multipliers:

\begin{theorem} In the distributional sense, we have
\begin{equation} \label{rel:fourier} \widehat{k^\beta}(\xi)=|\xi|_2^{-\beta}. \end{equation}
\end{theorem}
 We refer the reader to Section 3 of \cite{Haran2} for detailed proofs of such results. It rests on two ideas~: in the one hand holomorphic properties of the map $\beta \to k^\beta$ and in the other hand the direct calculus for $\Re (\beta)>0$ 
\begin{align*} 
\int_\QQ e^{2i\pi x\xi} \frac{d\mu(x)}{|x|_2^{-\beta+1}} & = |\xi|_2^{-\beta} \int_\QQ e^{2i\pi y} \frac{d\mu(y)}{|y|_2^{-\beta+1}} 
\\
&= |\xi|_2^{-\beta}  \sum_{k=-\infty}^{+\infty} 2^{k(\beta-1)}\int _{\abs{y} _2 = 2^k}e^{2i\pi y} d\mu(y) 
\\
 & = |\xi|_2^{-\beta} \left[\sum_{k\leq 0} 2^{k-1} 2^{k(\beta-1)} - 2^{\beta-1} \right] 
 \quad \hbox{(thanks to~(\ref{eq:intspheres}))}\\
 & = |\xi|_2^{-\beta} \left[\frac{1}{2(1-2^{-\beta})} -\frac{2^\beta}{2}\right]  
  = |\xi|_2^{-\beta} \frac{1-2^{\beta-1}}{1-2^{-\beta}} = |\xi|_2^{-\beta} \frac{\zeta(\beta)}{\zeta(1-\beta)}. 
\end{align*}

\begin{remark} By analogy with the Euclidean case, we can define a positive Laplacian operator $\Delta$ on $\QQ$ via the frequency space, as follows:
$$ \widehat{\Delta f} (\xi) := |\xi|_2^2 \widehat{f} (\xi).$$
Then, we emphasize that the Riesz operators can be also considered as a power of the Laplacian: ${\mathcal R}^\beta = \Delta^{-\beta/2}$.
We refer the reader to \cite{AKS} for more details about  $p$-adic pseudo-differential theory. 
\end{remark}

\noindent We deduce also the following properties (which can be obtained by a direct approach, see \cite{Haran}):

\begin{corollary} The Riesz operators satisfy to the semigroup property (also known as  ``Riesz reproduction formula'')~: for exponents $\beta,\beta'$ such that $\Re(\beta+\beta')<1$ \footnote{In \cite{AKS}, this assumption is not required and the author get the same properties in a distributional sense for every complex numbers $\beta,\beta'$}
$$ {\mathcal R}^\beta {\mathcal R}^{\beta'}= {\mathcal R}^{\beta+\beta'}$$
and in particular
$( {\mathcal R}^\beta)^{-1} = {\mathcal R}^{\beta}.$
\end{corollary}
 
\noindent Using the notion of Fourier series (developed in Section \ref{sec:fourier}), we can define multipliers on $\ZZ$ as follows:
\begin{definition} For $\beta>0$, we define on $\ZZ$
$$ \tilde k^\beta(x):=\frac{2}{\zeta(\beta)} |x|_2^{\beta-1}$$
and for $\beta<0$
$$ \tilde k^\beta(x):=p.v. \ \frac{2}{\zeta(-\beta)} |x|_2^{\beta-1}.$$
We write $\widetilde {\mathcal R} ^\beta$ the multiplier operator on $\ZZ$ defined as the convolution (on $\ZZ$) operator by $\tilde k^\beta$.
\end{definition}

\begin{proposition} We have the following Fourier representation: for all $\lambda\in \Lambda$
\begin{equation} \label{rel:fourier2} \F(\tilde k^\beta)(\lambda)=|\lambda|_2^{-\beta}. \end{equation}
\end{proposition}

\dem By definition, we have
$$ \F[\tilde k^\beta](\lambda) = \widehat{k^\beta{\bf 1}_{\ZZ}}(\lambda).$$
Then (\ref{rel:fourier2}) follows from the proof of (\ref{rel:fourier}).
\findem

\begin{corollary} \label{cor:Riesz} We deduce also that for exponents $\beta,\beta'$ with $\Re(\beta+\beta')<1$~:
$$ \widetilde {\mathcal R}^\beta \widetilde {\mathcal R}^{\beta'}=\widetilde {\mathcal R}^{\beta+\beta'},$$
in particular 
$(\widetilde {\mathcal R}^\beta)^{-1} = \widetilde {\mathcal R}^{\beta},$
where all these operators are defined on $\ZZ$.
\end{corollary}

 \section*{Acknowledgements} 
 We would like to thank Albert Cohen for his suggestions and fruitful remarks, and  
  Pierre Colmez for bringing to our attention the natural identification between the set of ends of the tree and the ring of 2-adic integers.
 






\begin{thebibliography}{99}

\bibitem{AKS} 
\newblock S. Albeverio, A. Yu. Khrennikov and V. M. Shelkovich,
\newblock \emph{Harmonic Analysis in the $p$-Adic Lizorkin Spaces: Fractional Operators, Pseudo-Differential Equations, $p$-Adic Wavelets, Tauberian Theorems},
\newblock J. Four. Anal. and Appl. \textbf{12} (2006), no. 4, 393--425.

\bibitem{baf} 
\newblock L. Baffico, C. Grandmont, Y. Maday and   A. Osses,
\newblock \emph{Homogenization of elastic media with gaseous inclusions}, 
\newblock Multiscale Modeling and Simulation, {\bf 7} (2008), 432--465.

\bibitem{cohen} 
\newblock A. Cohen, 
\newblock \emph{Numerical analysis of wavelet methods}, 
\newblock Studies in Mathematics and its Applications, {\bf 32}, 2003.


\bibitem{colmez}
\newblock P. Colmez, 
\newblock ``\'El\'ements d'analyse et d'alg\`ebre
(et de th\'eorie des nombres)'',
\newblock \'Editions de l'\'Ecole Polytechnique, 2009.



\bibitem{GMM} 
\newblock C. Grandmont, B. Maury and  N. Meunier, 
\newblock \emph{A viscoelastic model with non-local damping Application to the human lungs},
\newblock ESAIM : Mathematical Modelling and Numerical Analysis, {\bf 40} (2006),
201--224.

\bibitem{Haran} 
\newblock S. Haran, 
\newblock \emph{Riesz potentials and explicit sums in arithmetic}, 
\newblock Invent. math., {\bf 101} (1990), 697--703.

\bibitem{Haran2} 
\newblock S. Haran, 
\newblock \emph{Analytic potential theory over the $p$-adics}, 
\newblock Ann. Inst. Fourier, {\bf 43} (1993), no. 4, 905--944.

\bibitem{MMSV} 
\newblock B. Maury, N. Meunier, A. Soualah and L. Vial, 
\newblock \emph{Outlet dissipative conditions for air flow in the bronchial tree}, 
\newblock ESAIM : Proceedings,  {\bf 7} (2005), 1--10.

\bibitem{msv} 
\newblock B. Maury, D. Salort and C. Vannier,
\newblock \emph{Trace theorems for trees and application to the human lung}, 
\newblock Network and Heterogeneous Media {\bf 4} (2009), no. 3, 469--500.

\bibitem{soardi} 
\newblock P. M. Soardi,
\newblock \emph{Potential Theory on Infinite Networks}, 
\newblock Springer-Verlag, 1994.


\bibitem{weibelgeom}
\newblock E.R. Weibel,  
\newblock
in ``The Lung: Scientific Foundations'',  2nd edn Vol. 1, Eds Crystal, R. G.,West, J. B.,Weibel,
E. R. \& Barnes, P. J.,   Lippincott-Raven, Philadelphia, PA, 1997.


\end{thebibliography}
\end{document}